\documentclass[aps,tightenlines,nofootinbib,superscriptaddress,byrevtex]{revtex4}
\usepackage{amsmath,amssymb}
\usepackage{graphicx}
\usepackage{comment}

\def\str{{\mathrm{str}}}

\begin{document}

\unitlength=1mm

\def\a{{\alpha}}
\def\b{{\beta}}
\def\d{{\delta}}
\def\D{{\Delta}}
\def\e{{\epsilon}}
\def\g{{\gamma}}
\def\G{{\Gamma}}
\def\k{{\kappa}}
\def\l{{\lambda}}
\def\L{{\Lambda}}
\def\m{{\mu}}
\def\n{{\nu}}
\def\w{{\omega}}
\def\O{{\Omega}}
\def\S{{\Sigma}}
\def\s{{\sigma}}
\def\t{{\tau}}
\def\th{{\theta}}
\def\x{{\xi}}

\def\ol#1{{\overline{#1}}}

\def\Dslash{D\hskip-0.65em /}
\def\dslash{{\partial\hskip-0.5em /}}
\def\vslash{{\rlap \slash v}}
\def\qbar{{\overline q}}

\def\CPT{{$\chi$PT}}
\def\QCPT{{Q$\chi$PT}}
\def\PQCPT{{PQ$\chi$PT}}
\def\tr{\text{tr}}
\def\str{\text{str}}
\def\diag{\text{diag}}
\def\order{{\mathcal O}}
\def\vit{{\it v}}
\def\vD{\vit\cdot D}
\def\am{\alpha_M}
\def\bm{\beta_M}
\def\gm{\gamma_M}
\def\smb{\sigma_M}
\def\smt{\overline{\sigma}_M}
\def\tb{{\tilde b}}

\def\mc#1{{\mathcal #1}}

\def\Bbar{\overline{B}}
\def\Tbar{\overline{T}}
\def\cBbar{\overline{\cal B}}
\def\cTbar{\overline{\cal T}}
\def\pq{(PQ)}

\def\eqref#1{{(\ref{#1})}}

\newcount\hour \newcount\hourminute \newcount\minute 
\hour=\time \divide \hour by 60
\hourminute=\hour \multiply \hourminute by 60
\minute=\time \advance \minute by -\hourminute
\newcommand{\mydate}{\ \today \ - \number\hour :\number\minute}

\title{\textbf{Mixed Action Effective Field Theory: an Addendum}}

\author{Jiunn-Wei Chen}
\affiliation{Department of Physics and Center for Theoretical Sciences, National Taiwan University, Taipei 10617, Taiwan}

\author{Maarten Golterman}
\affiliation{Department of Physics and Astronomy, San Francisco State University, San Francisco, CA 94132, USA}

\author{Donal O'Connell}
\affiliation{School of Natural Sciences, Institute for Advanced Study, Princeton, NJ 08540, USA}

\author{ Andr\'e Walker-Loud} 
\affiliation{Department of Physics, College of William and Mary, Williamsburg, VA 23187-8795, USA}

\begin{abstract}
We correct a mistake in the literature regarding the additive lattice spacing corrections to the mixed valence-sea meson mass and discuss the consequences for mixed action extrapolation formulae.
\end{abstract}

\maketitle

\section{Mixed valence-sea mesons}
In the literature, it is well known that in mixed-action theories with exactly chiral (modulo
mass terms) valence
quarks there is one additional operator in the Symanzik expansion at order $a^2$ which arises from the breaking of the valence-sea symmetry~\cite{Bar:2003mh,Bar:2005tu}.  At the level of the chiral Lagrangian, this operator takes the form,%
\footnote{For all nonexplained notation, we refer to Ref.~\cite{Chen:2007ug}.  Further, the unitarity violations present in mixed action theories are the same, in spirit at least, as those found in partially quenched theories.  In partially quenched theories, unitarity is recovered in the limit the sea and valence quarks are degenerate.  For mixed action theories, one must also take the continuum limit to recover unitarity.  For a recent review of partially quenched theories, we refer the reader to Ref.~\cite{Sharpe:2006pu}.}
\begin{equation}\label{eq:VMix}
	\mc{L}_\mathrm{Mix} = -a^2 C_\mathrm{Mix} \str \left( T_3 \S T_3 \S^\dagger \right)
=-4\,a^2 C_\mathrm{Mix}\ \str \left( \mc{P}_S \S \mc{P}_S\S^\dagger \right) \, ,
\end{equation}
with
\begin{equation}\label{eq:T3}
	T_3 = \mc{P}_S - \mc{P}_V\, ,
\end{equation}
where $\mc{P}_S$ and $\mc{P}_V$ are sea and valence (ghost) projectors respectively.
  However, it is also often asserted that this is the only lattice-spacing dependent operator which contributes to the mass of a mixed valence sea meson at this order~\cite{Bar:2005tu,Tiburzi:2005is,Chen:2005ab,Bunton:2006va,Aubin:2006hg,Chen:2006wf,Chen:2007ug}.
In this note, we point out that this assertion is not correct, and discuss the consequences.  Before proceeding, we first revisit the proof of Ref.~\cite{Chen:2007ug}, that the mixed potential term, when restricted to one-loop
contributions, functions identically to a mixed valence-sea meson mass operator.
The key observation here is that there can only be one sea quark in the loop, as is
clear from the quark-flow picture for such loop contributions. 

This implies that if we expand out $\S=\mbox{exp}(2i\phi/f)$ in
terms of $\phi$, we will only use those terms in the expansion of Eq.~\eqref{eq:VMix} in which two of the indices on (two separate) $\phi$ fields correspond
to sea quarks, with the rest of the indices corresponding to valence quarks.  Because of the
$\mc{P}_S$ projectors in Eq.~\eqref{eq:VMix}, such terms only arise when we set either
$\S$ or $\S^\dagger$ equal to one.
Applying this rule immediately reduces Eq.~\eqref{eq:VMix} to (for a somewhat more explicit
argument, see Ref.~\cite{Chen:2007ug})
\begin{equation}\label{eq:CMix2}
	-a^2 C_\mathrm{Mix} \str \left( T_3 \S T_3 \S^\dagger \right) \bigg|^{2\phi_{vs}}_{(2N-2)\phi_{vv}}=
	-4\, a^2 C_\mathrm{Mix}\ \str \left[ \mc{P}_S \left( \S +\S^\dagger \right) \right]\, ,
\end{equation}
where the notation indicates that we restrict the term of order $\phi^{2N}$ in the expansion
of the operator on the left-hand side to have only two $\phi$ fields with one sea-quark
index each, whereas all the other indices are valence.

If $\mc{L}_\mathrm{Mix}$ were the only order $a^2$ term in the chiral lagrangian, the mixed meson
mass would be given by
\begin{equation}\label{eq:mixed_meson}
m_{vs}^2=B_0(m_{Q_v}+m_{Q_s})+a^2 \D_\mathrm{Mix}\ ,
\end{equation}
with
\begin{equation}\label{eq:C_Mix_Ratio2}
a^2 \D_\mathrm{Mix}=
\frac{16a^2C_\mathrm{Mix}}{f^2}\ .
\end{equation}
We may now compare the right-hand side of Eq.~(\ref{eq:CMix2}) with the 
 leading quark mass operator, $\mc{L}_{m_Q} = -(f^2B_0/4) \str \left[ m_Q (\S + \S^\dagger) \right]$, with the same flavor projection. It follows that the effects of this operator act at one-loop simply to shift the mixed valence-sea meson masses in all vertices and propagators.  It is important to note that the
replacement \eqref{eq:CMix2} {\em only} works for the goal of calculating one-loop corrections to valence quantities; $\mc{L}_\mathrm{Mix}$ does not contribute to the leading-order mass of a meson made out of sea quarks only, and the one-loop correction from $\mc{L}_\mathrm{Mix}$ to 
for instance, the mixed valence-sea meson mass breaks this rule~\cite{Orginos:2007tw}.

\subsection{Wilson sea fermions}
First consider a theory with chirally symmetric valence fermions and
$\mc{O}(a)$ improved Wilson sea fermions.  For an $\mc{O}(a)$ improved sea action, the leading terms in the lattice spacing dependent potential are ($\hat{a}\equiv 2W_0a$)~\footnote{For notation,
see Ref.~\cite{Bar:2003mh}.}
\begin{align}
\label{eq:WilsonSea}
\mc{L}_\mathrm{sea} &= -\hat{a}^2 \mc{V}_\mathrm{sea}\, ,\qquad \textrm{with}
\nonumber\\
\mc{V}_\mathrm{sea} &= W_6^\prime \left[ \str \left( \mc{P}_S \S + \mc{P}_S \S^\dagger \right) \right]^2
	+W_7^\prime \left[ \str \left( \mc{P}_S \S - \mc{P}_S \S^\dagger \right) \right]^2
	+W_8^\prime \str \left( \mc{P}_S \S \mc{P}_S \S + \mc{P}_S \S^\dagger \mc{P}_S \S^\dagger \right)\, ,
\end{align}
where again $\mc{P}_S$ is the sea projector.  Now the same argument used
above leads to the rule that either $\S$ or $\S^\dagger$ has to be replaced
by one, in all possible ways (two for each term, because all terms in
$\mc{V}_{sea}$ are quadratic in $\S$ and/or $\S^\dagger$).
This brings $\mc{V}_{sea}$ into the form
\begin{equation}\label{eq:WilsonSeaPot}
\hat{a}^2 \mc{V}_{sea} \bigg|^{2\phi_{vs}}_{(2N-2)\phi_{vv}}=
	 \left( 4N_s \hat{a}^2 W_6^\prime 
	+2\hat{a}^2 W_8^\prime \right) \str \left[ \mc{P}_S \left( \S + \S^\dagger \right) \right]\, .
\end{equation}
One immediately sees that, with this particular projection, the chiral structure of this operator is again identical to the mixed operator, Eq.~\eqref{eq:CMix2}, and therefore it will also act at one-loop as a mixed valence-sea meson mass shift in all vertices and propagators (in addition to shifting the sea-sea meson masses).  Thus, 
we now get another contribution to the mixed meson mass:
\begin{equation}
\label{eq:mixed_meson2}
m_{vs}^2= B_0(m_{Q_v}+m_{Q_s}) + a^2 \D_\mathrm{Mix} + a^2 \D^\prime_\mathrm{Mix}\, ,
\end{equation}
where %
\footnote{In the case of a twisted-mass sea, $a^2\D^\prime_\mathrm{Mix} \longrightarrow a^2\D^\prime_\mathrm{Mix} \cos^2 \w$, vanishing at maximal twist.  See Eq.~(A.12) of Ref.~\cite{Chen:2007ug}.} 
\begin{equation}
	a^2 \D^\prime_\mathrm{Mix} = \frac{8\hat{a}^2 (2N_s W_6^\prime +W_8^\prime)}{f^2}\, .
\end{equation}
We emphasize that the equality \eqref{eq:WilsonSeaPot} only holds under the specific
projection indicated in that equation.  For example, 
to leading order, the pure sea meson mass follows by direct calculation from Eq.~(\ref{eq:WilsonSea}), and is given by
\begin{equation}\label{eq:mixedmeson}
	m^2_{ss}=2B_0m_{Q_s}+a^2 \D_\mathrm{Sea}\ , \quad\textrm{ with}\quad 
	\D_\mathrm{Sea}=\frac{32\hat{a}^2}{f^2}\left(W_8'+ N_s W_6'\right)\ .
\end{equation}
The $W_7^\prime$ operator does not contribute to the meson masses at this order, but it does give rise to a lattice spacing dependent hairpin interaction~\cite{Golterman:2005xa}.  However, as discussed in Ref.~\cite{Golterman:2005xa}, this new hairpin interaction has the same structure as the hairpins arising from the sea-valence quark mass difference.  Consequently, the effects of this new hairpin can be completely absorbed as a shift of the partial quenching parameters, $\D_{PQ}^2 = m_{ss}^2 - m_{vv}^2$.  This is crucial to the arguments in Refs.~\cite{Chen:2006wf,Chen:2007ug} regarding the universal nature of mixed action effective field theories with chirally symmetric valence fermions.  With this shift, the partial quenching parameter becomes~\cite{Golterman:2005xa,Chen:2007ug}
\begin{equation}
	\D_{PQ}^2 \longrightarrow m_{ss}^2 -m_{vv}^2 + \hat{a}^2 \g_{ss} N_s\, ,
	\quad\textrm{with}\quad
	\g_{ss} = \frac{32 W_7^\prime}{f^2}\, .
\end{equation}
One should caution however, that a determination of $\g_{ss}$ will be difficult as it is a discretization correction to the mass of the $\eta^\prime$.

\subsection{Staggered sea fermions}
A similar analysis applies to the case that we choose the sea quarks to be staggered.
The sea potential for staggered fermions is~\cite{Lee:1999zxa,Aubin:2003mg,Bar:2005tu}%
\footnote{For notation,
see Ref.~\cite{Bar:2005tu}.}
\begin{equation}
	\mc{L}_\mathrm{sea} = -a^2 \mc{U}_S -a^2 \mc{U}^\prime_S\, ,
\end{equation}
with 
\begin{align}
\label{eq:Us}
\mc{U}_S =&\ C_1 \str \left( \hat{\xi}_5 \mc{P}_S \S \hat{\xi}_5 \mc{P}_S \S^\dagger \right)
	+\frac{C_3}{2}\sum_\nu \str \left( \hat{\xi}_\nu \mc{P}_S \S \hat{\xi}_\nu \mc{P}_S \S + \textrm{h.c.} \right)
\nonumber\\
	&+\frac{C_4}{2}\sum_\nu \str \left( \hat{\xi}_{\nu5} \mc{P}_S \S \hat{\xi}_{5\nu} \mc{P}_S \S + \textrm{h.c.} \right)
	+C_6 \sum_{\mu < \nu} \str \left( \hat{\xi}_{\mu\nu}\mc{P}_S \S \hat{\xi}_{\nu\mu}\mc{P}_S \S^\dagger \right)\, ,
\nonumber\\
\mc{U}^\prime_S =&\ \frac{C_{2V}}{4} \sum_\nu \left[ \str(\hat{\xi}_\nu \mc{P}_S \S) \str( \hat{\xi}_\nu \mc{P}_S \S) + \textrm{h.c.} \right]
	+\frac{C_{2A}}{4} \sum_\nu \left[ \str(\hat{\xi}_{\nu5} \mc{P}_S \S) \str( \hat{\xi}_{5\nu} \mc{P}_S \S) + \textrm{h.c.} \right]
	\nonumber\\&
	+\frac{C_{5V}}{2}\sum_\nu \str \left(\hat{\xi}_\nu \mc{P}_S \S \right) 
		\str \left( \hat{\xi}_\nu \mc{P}_S \S^\dagger \right)
	+\frac{C_{5A}}{2}\sum_\nu \str \left(\hat{\xi}_{\nu5} \mc{P}_S \S \right) 
		\str \left( \hat{\xi}_{5\nu} \mc{P}_S \S^\dagger \right)\, .
\end{align}
Applying the same argument as before, we find that now 
\begin{equation}\label{eq:StagSeaPot}
a^2 \mc{U}_{S} + a^2 \mc{U}^\prime_S \bigg|^{2\phi_{vs}}_{(2N-2)\phi_{vv}}
	= a^2\left( C_1 +4C_3 +4C_4 +6C_6 \right) \str \left[ \mc{P}_S \left( \S + \S^\dagger \right) \right]\, .
\end{equation}
Again, with this restriction, these operators have the same chiral structure as the mixed term, Eq.~\eqref{eq:CMix2}, and therefore we find again a mixed meson mass of the form (\ref{eq:mixed_meson2}), but now with 
\begin{align}
\label{eq:Stagmixprime}
	a^2 \D^\prime_\mathrm{Mix} &= \frac{4a^2 (C_1 +4C_3 + 4C_4 + 6C_6)}{f^2}
\nonumber\\
	&= \frac{1}{8}a^2\D_A^2 +\frac{3}{16}a^2\D_T^2
	+\frac{1}{8}a^2\D_V^2 + \frac{1}{32}a^2\D_I^2\, ,
\end{align}
where the latter expression is in terms of the taste splittings of Ref.~\cite{Aubin:2004fs}.  Note
that the $U(1)_\e$ symmetry of the staggered lattice action prohibits a $W_7^\prime$--like term.

\subsection{Generic sea fermions}

One might ask how general reductions of the type
(\ref{eq:WilsonSeaPot}) and (\ref{eq:StagSeaPot}) are, {\it i.e.}, whether
a similar argument would apply for any kind of sea quarks.  Whether this
is the case or not depends on whether the sea sector has enough symmetry.

In general, order $a^2$ corrections of the type of $\mc{L}_\mathrm{sea}$
come from four-fermion operators in the Symanzik effective
action. Following the general steps of the analysis of
Refs.~\cite{Bar:2003mh,Bar:2005tu}, such operators translate into
chiral perturbation theory operators which are bilinear in $\S$ and $\S^\dagger$. After our
reduction rule, in which one of the $\S$ or $\S^\dagger$ gets replaced by
one, these operators then all have to take the form $\str [\mc{P}_S (X\S+\S^\dagger
X^\dagger)] $, in which $X$ is some flavor matrix. In the Wilson and
staggered cases, we have that $X=1$, but if the sea sector has less
symmetry, a nontrivial $X$ may be possible. An example is provided
by Dirac-K\"ahler fermions \cite{Becher:1982ud}, which essentially
are staggered fermions without shift symmetry \cite{Golterman:1984cy}.
If one does not impose shift symmetry, many more four-fermion operators
are possible in the Symanzik theory \cite{Lee:1999zxa}, leading to
chiral perturbation theory operators of the form $\str(Y_1P_S\S Y_2P_S\S^\dagger)$ or $\str(Y_1P_S\S
Y_2P_S\S+\mathrm{h.c.})$ with $Y_1\ne Y_2$, unlike Eq.~(\ref{eq:Us}),
leading to $X\ne 1$. However, the same lack of symmetry also leads to
additive, taste-breaking quark mass renormalization \cite{Mitra:1983bi},
making these fermions less attractive for applications to lattice QCD.

We can understand this point somewhat more generally by considering the
hierarchy of symmetry violations that occur in the sea. In the continuum
theory, flavor symmetry SU($N_s$) is broken by quark mass differences. In
the lattice theory, SU($N_s$) is broken to some group $G'$ by lattice
spacing artifacts; this group $G'$ may be further broken to some group $G$
by quark mass differences. Physically, the nicest case occurs when $G'
\neq G$, and the action of $G'$ is via an $N_s$ dimensional irreducible
representation. In this case, even on the lattice, there is a notion of
a flavor symmetry acting on all the sea quarks; in addition, it is easy
to see that with these assumptions the flavor matrix $X$ we encountered
above must be proportional to the identity as we will now show. 

Our assumption is that there is some symmetry $G'$ of the theory
which is only broken by quark mass differences. Now, at order $a^2$,
we have encountered the operator $\mc{O} = \str [\mc{P}_S (X \Sigma +
\Sigma^\dagger X^\dagger)] $, describing lattice spacing effects; as such,
$\mc{O}$ is invariant under the larger group $G'$.  Suppose $U \in G'$
and consider rotating $\Sigma \rightarrow U^\dagger \Sigma U$. Since
this is a symmetry of $\mc{O}$, we find that $U X U^\dagger = X$ for all
$U \in G'$. By Schur's Lemma, it follows that $X$ is proportional to the
identity matrix. Note that this argument does not apply to Dirac-K\"ahler
fermions, because the lack of shift symmetry at the lattice level leads
to loop corrections which violate mass degeneracy~\cite{Mitra:1983bi}.
Staggered fermions do fit in, because for $N_f$ staggered fermions,
ignoring quark masses, there is an $SU(N_f)\times\Gamma_4$
symmetry, which acts irreducibly on the sea quarks.  Here $\Gamma_4$ is
the 32-element finite group generated by the matrices $\xi_\mu$; the
${\hat\xi}_\mu$ in Eq.~(\ref{eq:Us}) above are equal to ${\bf 1}\times\xi_\mu$,
with $\bf 1$ the $N_f\times N_f$ unit matrix.

\subsection{Consequences for mixed action extrapolation formulae}
The consequences of these previously neglected effects are in fact very mild.  In Ref.~\cite{Chen:2007ug}, it was demonstrated that the mixed potential operator, when projected onto two valence-sea mesons and $2N-2$ valence-valence mesons, has the chiral structure of Eq.~\eqref{eq:CMix2}, and consequently acts exactly like a mass term for the mixed mesons in all vertices and propagators through one-loop order.  Here we have shown that the sea potential under the same projection also has the same chiral structure as the mixed potential, Eqs.~\eqref{eq:WilsonSeaPot} and \eqref{eq:StagSeaPot} for the case of Wilson and staggered sea actions respectively.  Consequently, these operators, at the one loop level, will also act just like mixed meson mass operators in all vertices and propagators.  Importantly, the arguments of Refs.~\cite{Chen:2006wf,Chen:2007ug} regarding the universal nature of the mixed action chiral extrapolation formulae, and the vanishing of lattice-spacing dependent counter terms with the use of on-shell renormalization still hold.

Practically, the mixed meson mass renormalization calculated in Refs.~\cite{Orginos:2007tw,Aubin:2008ie} for the specific mixed action of domain-wall~\cite{Kaplan:1992bt,Shamir:1993zy,Furman:1994ky} valence fermions and the asqtad improved~\cite{Orginos:1998ue,Orginos:1999cr} MILC staggered sea fermions~\cite{Bernard:2001av}, was not of the parameter $a^2\D_\mathrm{Mix}$, but rather 
\begin{equation}
	m_{vs}^2 - \frac{1}{2}m_{vv}^2 - \frac{1}{2}m_{ss}^2 = a^2 \D_\mathrm{Mix}
	+\frac{1}{8}a^2\D_A^2 +\frac{3}{16}a^2\D_T^2	+\frac{1}{8}a^2\D_V^2 + \frac{1}{32}a^2\D_I^2\, .
\end{equation}
Here $m^2_{ss}=2B_0m_{Q_s}$ is the mass-squared of the exact staggered Goldstone boson,
made out of sea quarks only.
Using the known values of the staggered meson mass splittings~\cite{Aubin:2004fs}, and the calculated value of the mass splitting on the coarse ($a\sim 0.125~\texttt{fm}$) ensemble~\cite{Orginos:2007tw,Aubin:2008ie}, one finds
\begin{equation}
	a^2\D_\mathrm{Mix} \sim (160 \texttt{ MeV})^2\, .
\end{equation}
However, the total mixed meson mass shift is still given by the numerical shifts calculated in Refs.~\cite{Orginos:2007tw,Aubin:2008ie}, which is what is practically important for accounting for discretization effects in mixed action chiral extrapolations.


\begin{acknowledgments} 
We would like to thank Oliver B\"{a}r and Claude Bernard for useful correspondence.  JWC is supported by the National Science Council of Taiwan. DOC is the Martin A. and Helen Chooljian Member at the Institute for Advanced Study, and was supported in part by the US DOE under contract DE-FG02-90ER-40542.  MG is supported in part by the US DOE.  AWL is supported under the U.S. DOE OJI grant DE-FG02-07ER-41527.
\end{acknowledgments}

\bibliography{errata}

\end{document}